\documentclass[floatfix,reprint,superscriptaddress,showpacs,preprintnumbers,nofootinbib,amsmath,amssymb,epsfig,pre,floats,letterpaper]{revtex4-1}

\usepackage{amsmath}
\usepackage{color}
\usepackage{amssymb}
\usepackage{graphicx}
\usepackage{epstopdf}
\usepackage{hyperref}

\newcommand{\beq}{\begin{equation}}
\newcommand{\eeq}{\end{equation}}
\newcommand{\e}{\mathrm{e}}
\newcommand{\la}{\langle}
\newcommand{\ra}{\rangle}

\begin{document}

\title{A maximum entropy framework for non-exponential distributions}

\author{Jack Peterson}
\affiliation{Oregon State University, Department of Mathematics, Corvallis, OR}
\affiliation{Laufer Center for Physical and Quantitative Biology, Departments of Physics and Chemistry, Stony Brook University, NY}

\author{Purushottam D. Dixit}
\affiliation{Center for Computational Biology and Bioinformatics, Department of Biomedical Informatics, Columbia University, New York, NY}

\author{Ken A. Dill}
\affiliation{Laufer Center for Physical and Quantitative Biology, Departments of Physics and Chemistry, Stony Brook University, NY}

\begin{abstract}
Probability distributions having power-law tails are observed in a broad range of social, economic, and biological systems.  We describe here a potentially useful common framework.  We derive distribution functions $\{p_k\}$ for situations in which a `joiner particle' $k$ pays some form of price to enter a `community' of size $k-1$, where costs are subject to economies-of-scale (EOS).  Maximizing the Boltzmann-Gibbs-Shannon entropy subject to this energy-like constraint predicts a distribution having a power-law tail; it reduces to the Boltzmann distribution in the absence of EOS.  We show that the predicted function gives excellent fits to 13 different distribution functions, ranging from friendship links in social networks, to protein-protein interactions, to the severity of terrorist attacks.  This approach may give useful insights into when to expect power-law distributions in the natural and social sciences.
\end{abstract}

\maketitle

Probability distributions are often observed to have power-law tails, particularly in social, economic, and biological systems.  Examples include distributions of fluctuations in financial markets \cite{Mantegna_1995}, the populations of cities \cite{Zipf_1949}, the distribution of website links~\cite{Broder_2000}, and others~\cite{Newman_2005,Clauset_2009}.  Such distributions have generated much popular interest \cite{Taleb_2007,Bremmer_2009} because of their association with rare but consequential events, such as stock market bubbles and crashes.

If sufficient data is available, finding the mathematical shape of a distribution function can be as simple as curve-fitting, with a follow-up determination of the significance of the mathematical form used to fit it.  On the other hand, it is often interesting to know if the shape of a given distribution function can be explained by an underlying generative principle.  Principles underlying power-law distributions have been sought in various types of models.  For example, the power-law distributions of node connectivities in social networks have been derived from dynamical network evolution models~\cite{Vazquez_2003_Complexus,Berg_2004,maslov2009toolbox,pang2011toolbox,Leskovec_2010,karagiannis2010power,shou2011measuring, fortuna2011evolution,Peterson_2012,pang2013universal}.  A large and popular class of such models is based on the `preferential attachment' rule \cite{Simon_1955,Price_1976,Barabasi_1999,vazquez2003growing,yook2002modeling,capocci2006preferential,newman2001clustering,jeong2003measuring,poncela2008complex, Peterson_2010}, wherein it is assumed that new nodes attach preferentially to the largest of the existing nodes.  Explanations for power-laws are also given by Ising models in critical phenomena~\cite{Fisher_1974,Yeomans_1992,Stanley_1999,gefen1980critical,fisher1986scaling,suzuki1968dynamics,glauber1963time}, network models with thresholded `fitness' values \cite{Caldarelli_2002} and random-energy models of hydrophobic contacts in protein interaction networks \cite{Deeds_2006}.

However, such approaches are often based on particular mechanisms or processes.  They often predict particular power-law exponents, for example.  Our interest here is in finding a broader vantage point, as well as a common language, for describing a range of distributions, from power-law to exponential.  For deriving exponential distributions, a well-known general principle is the method of Maximum Entropy (Max Ent) in statistical physics~\cite{Shore_1980,jaynes1957information}. In such problems, you want to choose the best possible distribution from all candidate distributions that are consistent with certain set of constrained moments, such as the average energy.  For this type of problem, which is highly underdetermined, a principle is needed for selecting a `best' mathematical function from among alternative model distribution functions.  To find the mathematical form of the distribution function $p_k$ over states $k = 1, 2, 3, \ldots$, the Max Ent principle asserts that you should maximize the Boltzmann-Gibbs-Shannon (BGS) entropy functional $S[\{p_k\}] = -\sum_k p_k \log p_k$ subject to constraints, such as the known value of the average energy $\la E \ra$.  This procedure gives the exponential (Boltzmann) distribution, $p_k \propto \e^{-\beta E_k}$, where $\beta$ is the Lagrange multiplier that enforces the constraint.  This variational principle has been the subject of various historical justifications.  It is now commonly understood as the approach that chooses the least-biased model that is consistent with the known constraint(s)~\cite{Press2012}.

\begin{figure*}
\centerline{
\includegraphics[width=0.65\textwidth]{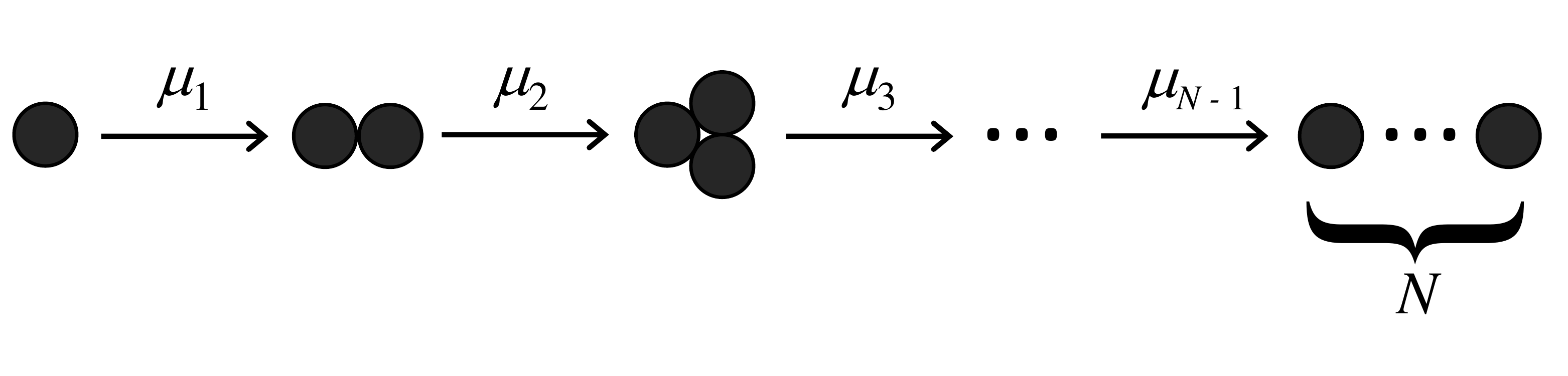}
}
\caption{$\mu_k$ is the \emph{joining cost} for a particle to join a size {$k-1$} community.  This diagram can describe particles forming colloidal clusters, or {social} processes such as people joining cities, citations added to papers, or {link creation} in a social network.}
\label{fig:schematic}
\end{figure*}

Is there an equally compelling principle that would select fat-tailed distributions, given limited information?  There is a large literature that explores this.  Inferring non-exponential distributions can be done by maximizing a different mathematical form of entropy, rather than the Boltzmann-Gibbs-Shannon form.  Examples of these non-traditional entropies include those of Tsallis~\cite{tsallis1988possible},  Renyi~\cite{rrnyi1961measures}, and others~\cite{aczel1975measures,amari1985differential}.  For example, the Tsallis entropy is defined as $\frac{K}{1-q} \left(\sum_{k}p_{k}^{q} -1\right),\label{tsallisent}$ where $K$ is a constant and $q$ is a parameter for the problem at hand.  Such methods otherwise follow the same strategy as above: maximizing the chosen form of entropy subject to an extensive energy constraint gives non-exponential distributions. The Tsallis entropy has been applied widely~\cite{lutz2003anomalous,douglas2006tunable,burlaga2005triangle,pickup2009generalized,devoe2009power,plastino1995non, tsallis1996anomalous,caruso2008nonadditive,abe2000axioms,gell2004nonextensive}.

However, we adopt an alternative way to infer non-exponential distributions.  To contrast our approach, we first switch from probabilities to their logarithms.  Logarithms of probabilities can be parsed into energy-like and entropy-like components, as is standard in statistical physics.  Said differently, a nonexponential distribution that is derived from a Max Ent principle requires that there be non-extensivity in either an energy-like or entropy-like term; that is, it is non-additive over independent subsystems, not scaling linearly with system size.   Tsallis and others have chosen to assign the non-extensivity to an entropy term, and retain extensivity in an energy term.  Here, instead, we keep the canonical BGS form of entropy, and invoke a non-extensive energy-like term.  In our view, only the latter approach is consistent with the principles elucidated by Shore and Johnson (SJ)~\cite{Shore_1980} (reviewed in~\cite{Press2012}).  Shore and Johnson (SJ) showed that the Boltzmann-Gibbs-Shannon form of entropy is uniquely the mathematical function that ensures satisfaction of the addition and multiplication rules of probability.  SJ asserts that any form of entropy other than BGS will impart a bias that is unwarranted by the data it aims to fit.  We regard the SJ argument as a compelling first-principles basis for defining a proper variational principle for modeling distribution functions.  Here, we describe a variational approach based on the BGS entropy function, and we seek an explanation for power-law distributions in the form of an energy-like function instead.

\section{Theory\label{sc:th}}

\subsection{Assembly of simple colloidal particles}

We frame our discussion in terms of a `joiner particle' that enters a cluster or community of particles, as shown in Fig.~\ref{fig:schematic}.  On the one hand, this is a natural way to describe the classical problem of the colloidal clustering of physical particles; it is readily shown (reviewed below) to give an exponential distribution of cluster sizes.  On the other hand, this general description also pertains more broadly, such as when people populate cities, links are added to websites, or when papers accumulate citations.  We want to compute the distribution, $p_k$, of populations of communities having size $k = 1, 2, \ldots, N$.

To begin, we express a cumulative `cost' of joining.  For particles in colloids, this cost is expressed as a chemical potential, i.e., a free energy per particle.  If $\mu_j$ represents the cost of adding particle $j$ to a cluster of size $j-1$, the cumulative cost of assembling a whole cluster of $k$ particles is the sum,
\beq
w_k = \sum_{j=1}^{k-1} \mu_j. \label{eq:cumu}
\eeq

Max Ent asserts that we should choose the probability distribution that has the maximum entropy amongst all candidate distributions that are consistent with the mean value $\la w \ra$ of the total cost of assembly~\cite{Dill_2010},
\beq\label{eq:sumQ}
p_k = \frac{\e^{-\lambda w_k}}{\displaystyle\sum_i \e^{-\lambda w_i}},
\eeq
where $\lambda$ is a Lagrange multiplier that enforces the constraint.

In situations where the cost of joining does not depend on the size of the community a particle joins, then $\mu_k = \mu^\circ$, where $\mu^\circ$ is a constant.  The cumulative cost of assembling the cluster is then
\beq
\label{eq:cumucost}
w_k = \left(k-1\right) \mu^\circ.
\eeq
Substituting into Eq.~\ref{eq:sumQ} and absorbing the Lagrange multiplier $\lambda$ into $\mu^\circ$ yields the grand canonical exponential distribution, well-known for this problems such as this:
\beq
p_k = \frac{\e^{-\mu^\circ k}}{\displaystyle\sum_i \e^{-\mu^\circ i}}.
\eeq

In short, when the joining cost of a particle entry is independent of the size of the community it enters, the community size distribution is exponential.

\subsection{Communal assemblies and `economies of scale'}

Now, we develop a general model of communal assembly based on `economies of scale'.  Consider a situation where the joining cost for a particle depends on the size of the community it joins.  In particular, consider situations in which the costs are lower for joining a larger community.  Said differently, the `cost-minus-benefit' function $\mu_k$ is now allowed to be subject to `economies of scale', which, as we note below, can also be interpreted instead as a form of discount in which the community pays down some of the joining costs for the joiner particle.

To see the idea of economy-of-scale cost function, imagine building a network of telephones.  In this case, a community of size $1$ is a single unconnected phone.  A community of size $2$ is two connected phones, etc.  Consider the first phone: The cost of creating the first phone is high because it requires initial investment in the phone assembly plant.  And the benefit is low, because there is no value in having a single phone.  Now, for the second phone, the cost-minus-benefit is lower.  The cost of producing the second phone is lower than the first since the production plant already exists.  And the benefit is higher because two connected phones are more useful than one unconnected phone.  For the third phone, the cost-minus-benefit is even lower than for the second because the production cost is even lower (economy of scale) and because the benefits increase with the number of phones in the network.

To illustrate, suppose the cost-minus-benefit for the first phone is $150$, for the second phone is $80$, and for the third phone is $50$.  To express these cost relationships, we define an `intrinsic cost' for the first phone (joiner particle), $150$ in this example.  And, we define the difference in cost-minus-benefit between the first and second phones as the discount provided `by the first phone' when the second phone `joins the community' of two phones.  In this example, the first phone provides a discount of $70$ when the second phone joins.  Similarly, the total discount provided by the two-phone community is $100$ when the third phone joins the community.

\begin{figure*}
\centerline{
\includegraphics[width=0.32\textwidth]{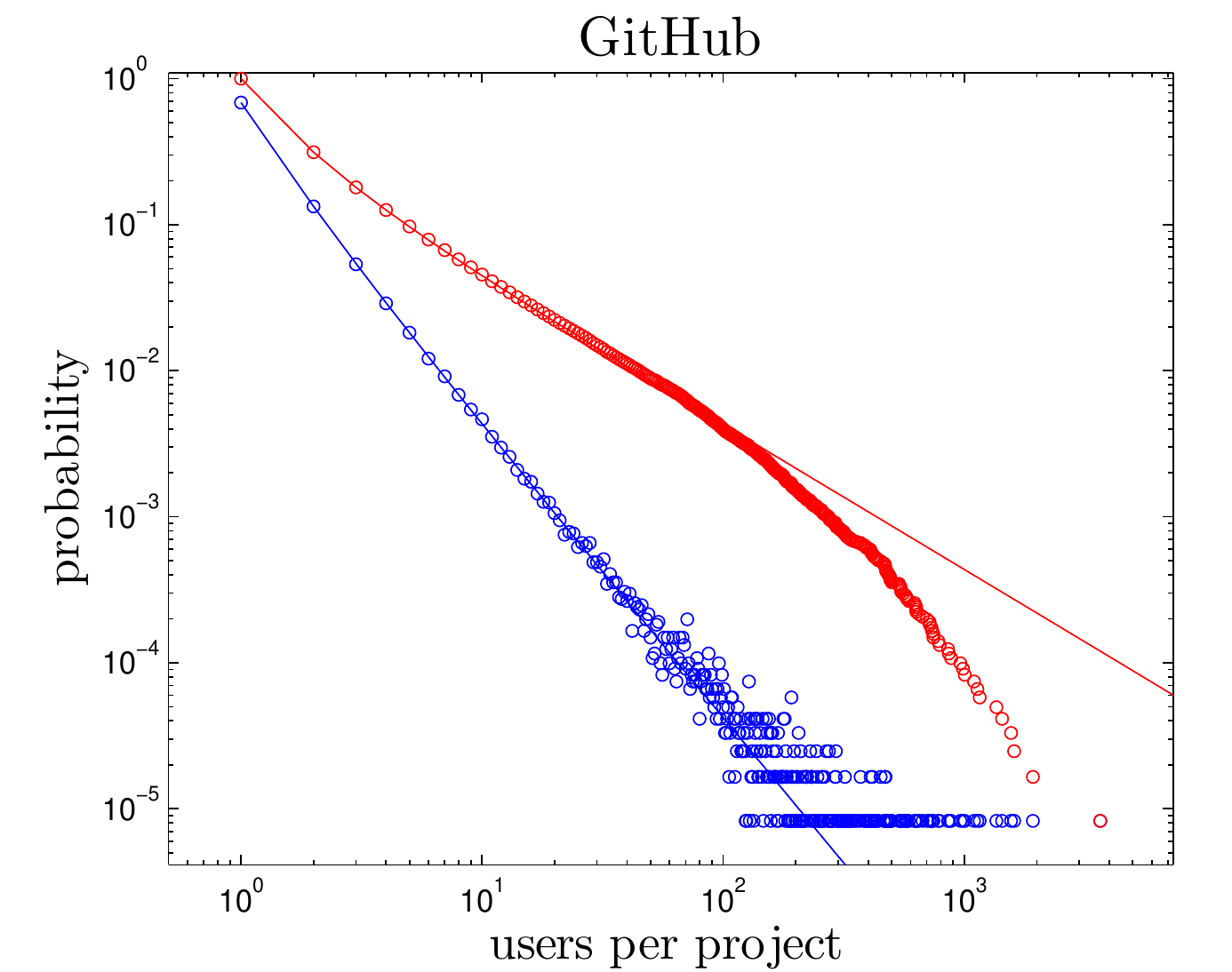}
\includegraphics[width=0.32\textwidth]{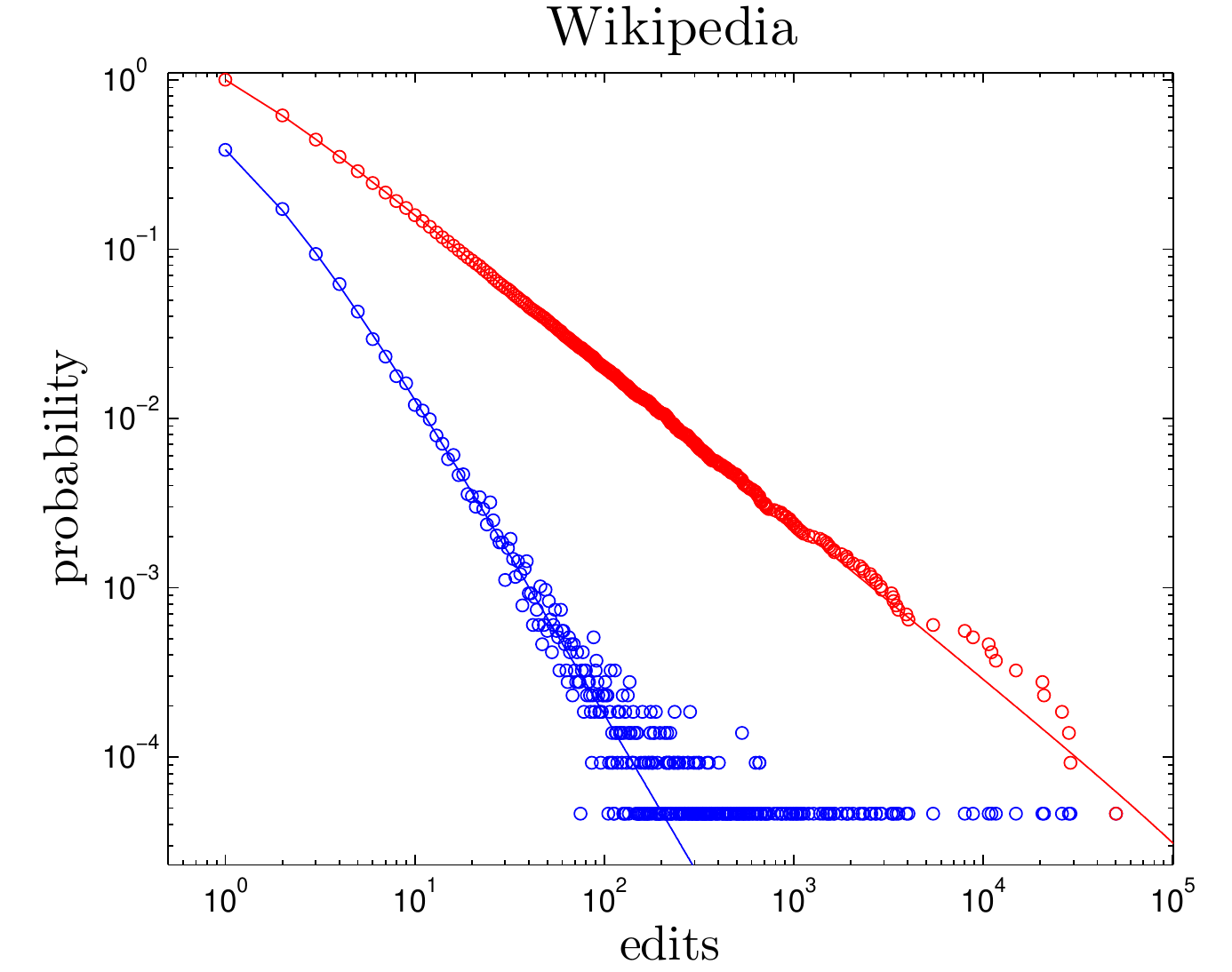}
\includegraphics[width=0.32\textwidth]{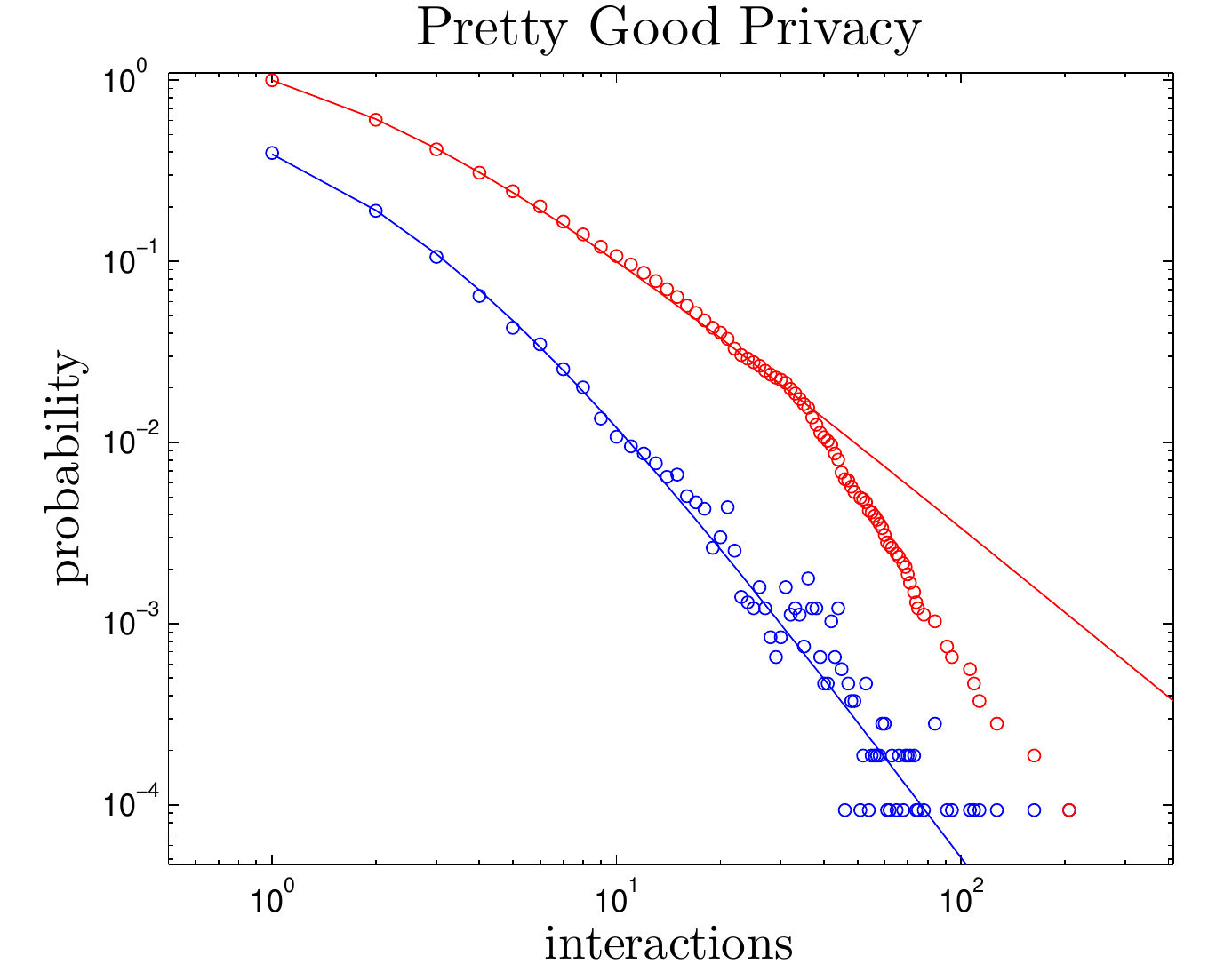}
}
\centerline{
\includegraphics[width=0.32\textwidth]{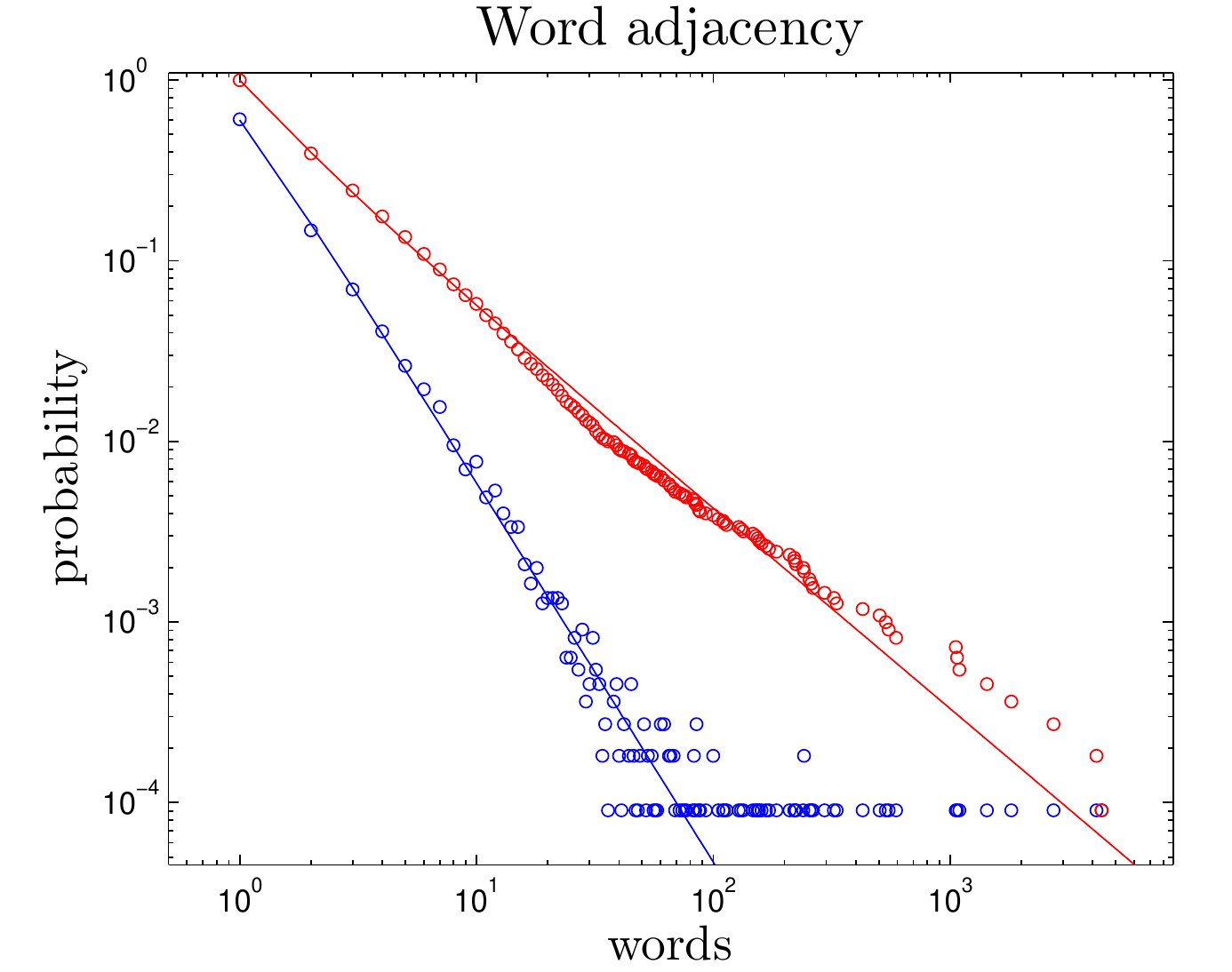}
\includegraphics[width=0.32\textwidth]{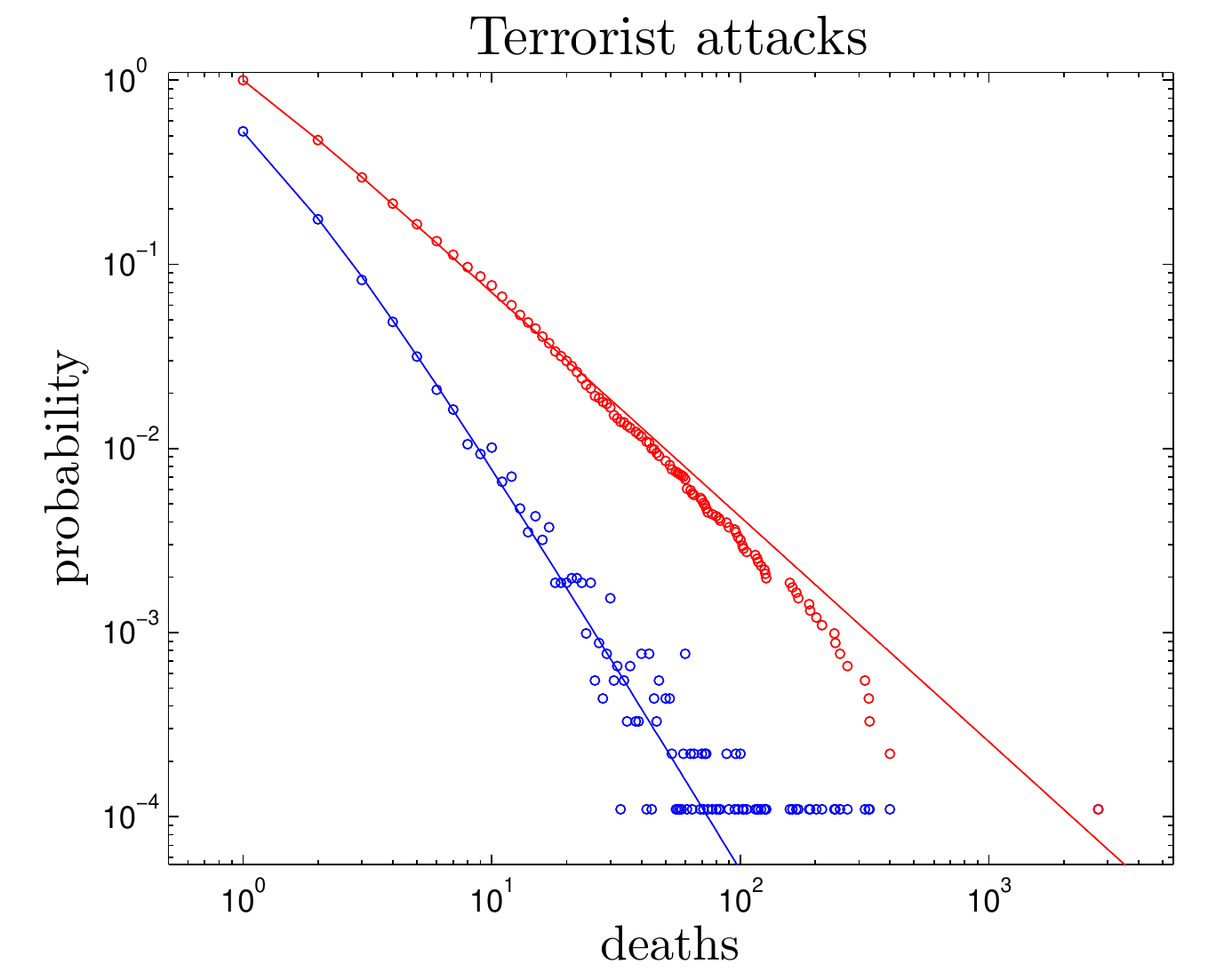}
\includegraphics[width=0.32\textwidth]{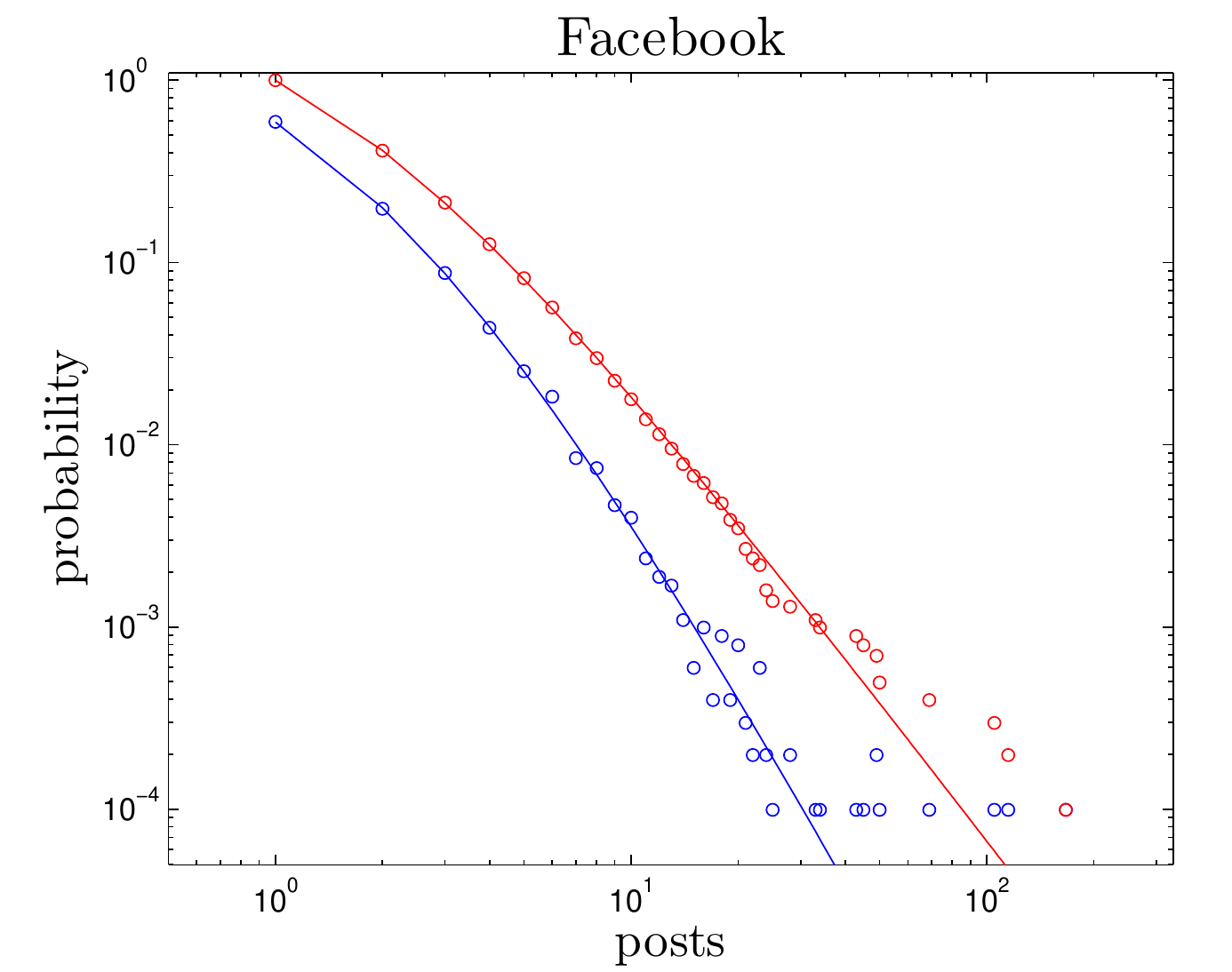}
}
\centerline{
\includegraphics[width=0.32\textwidth]{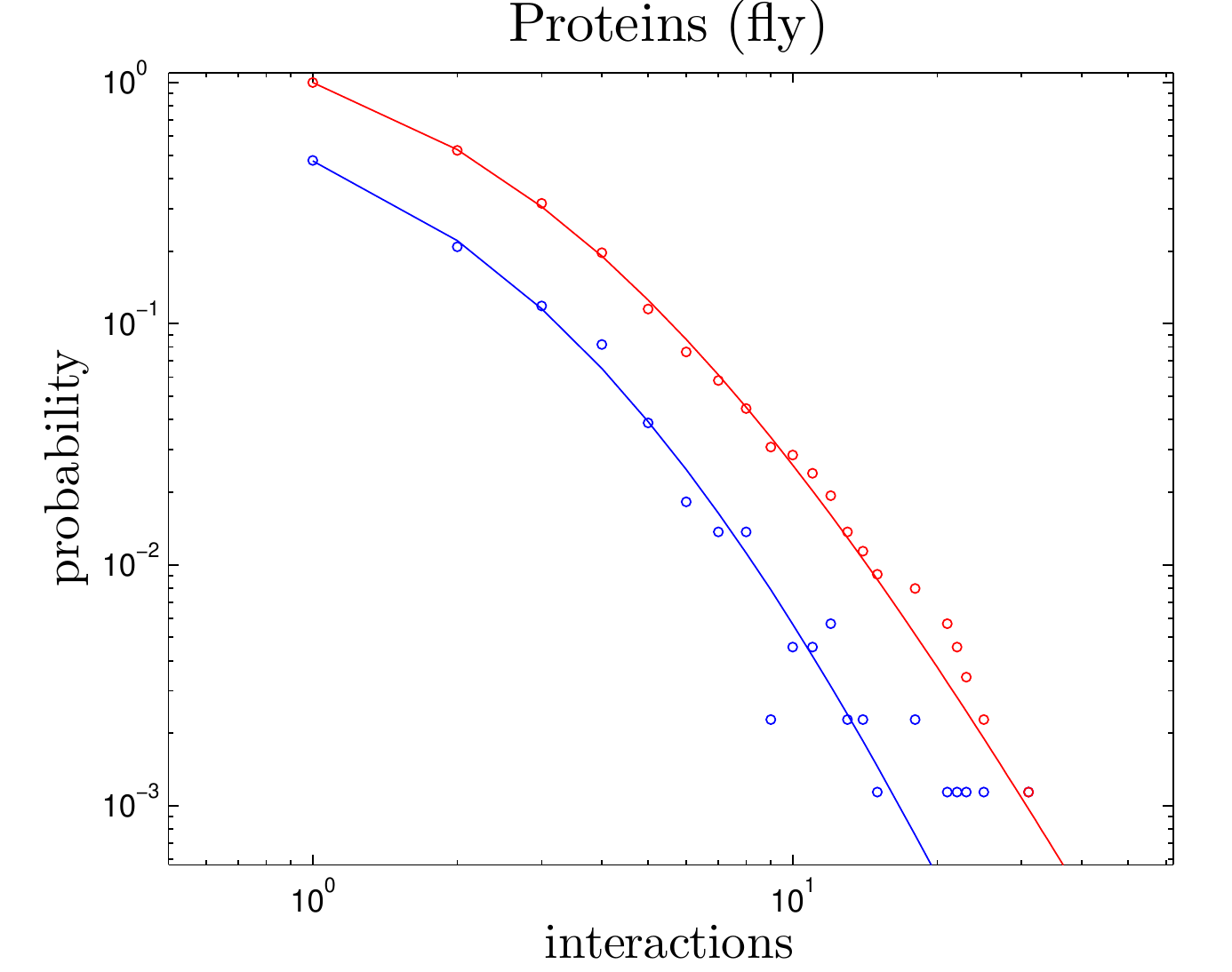}
\includegraphics[width=0.32\textwidth]{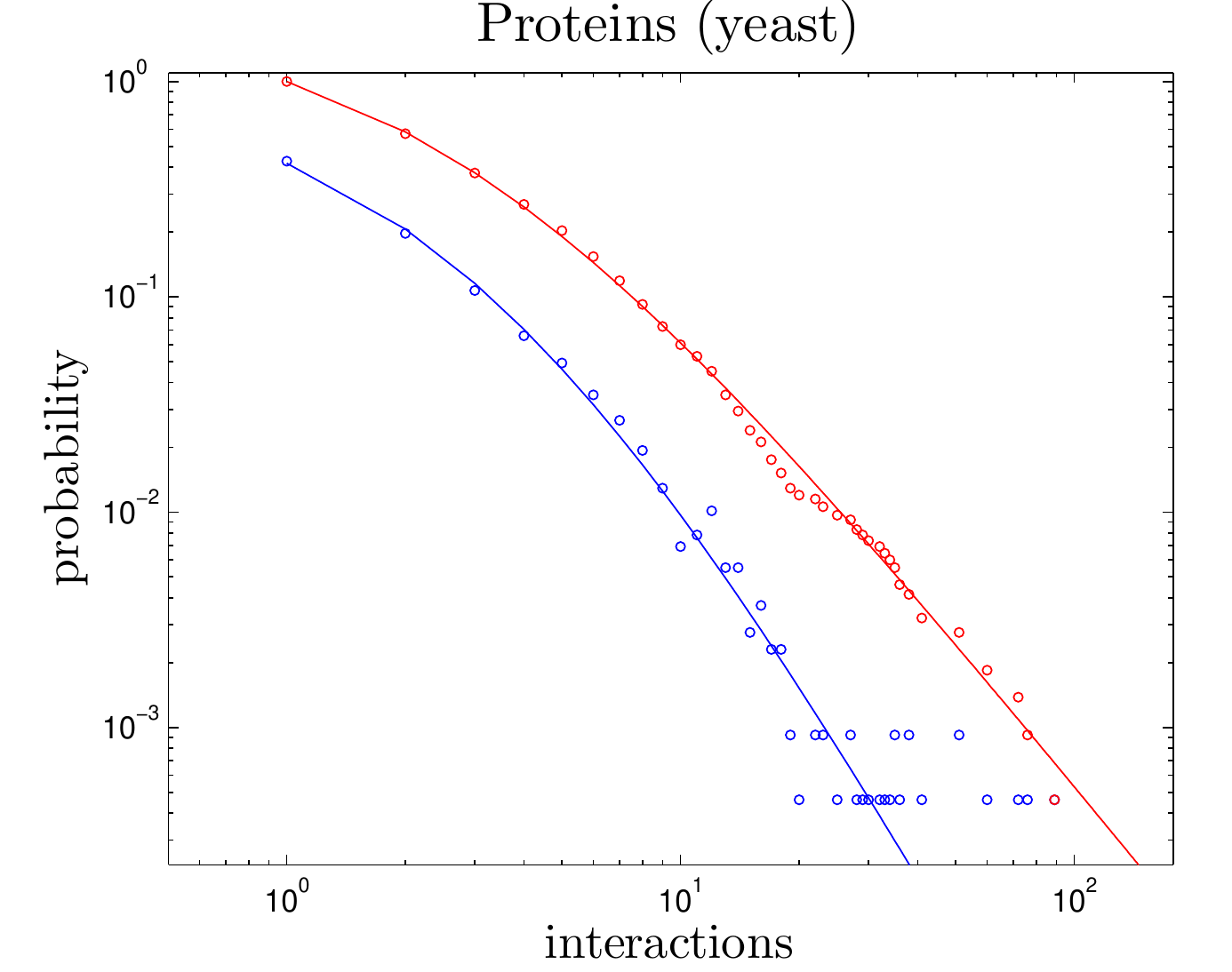}
\includegraphics[width=0.32\textwidth]{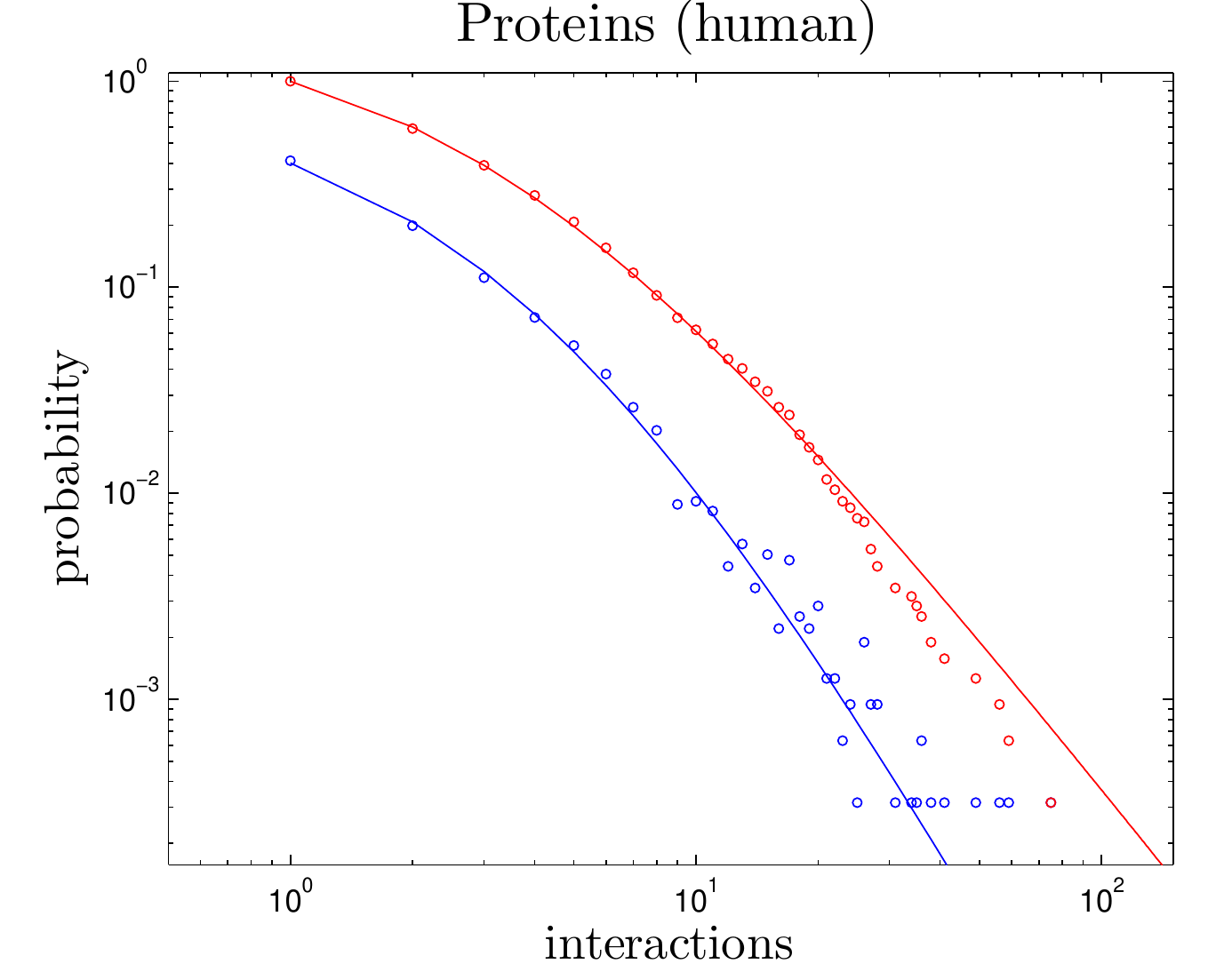}
}
\centerline{
\includegraphics[width=0.32\textwidth]{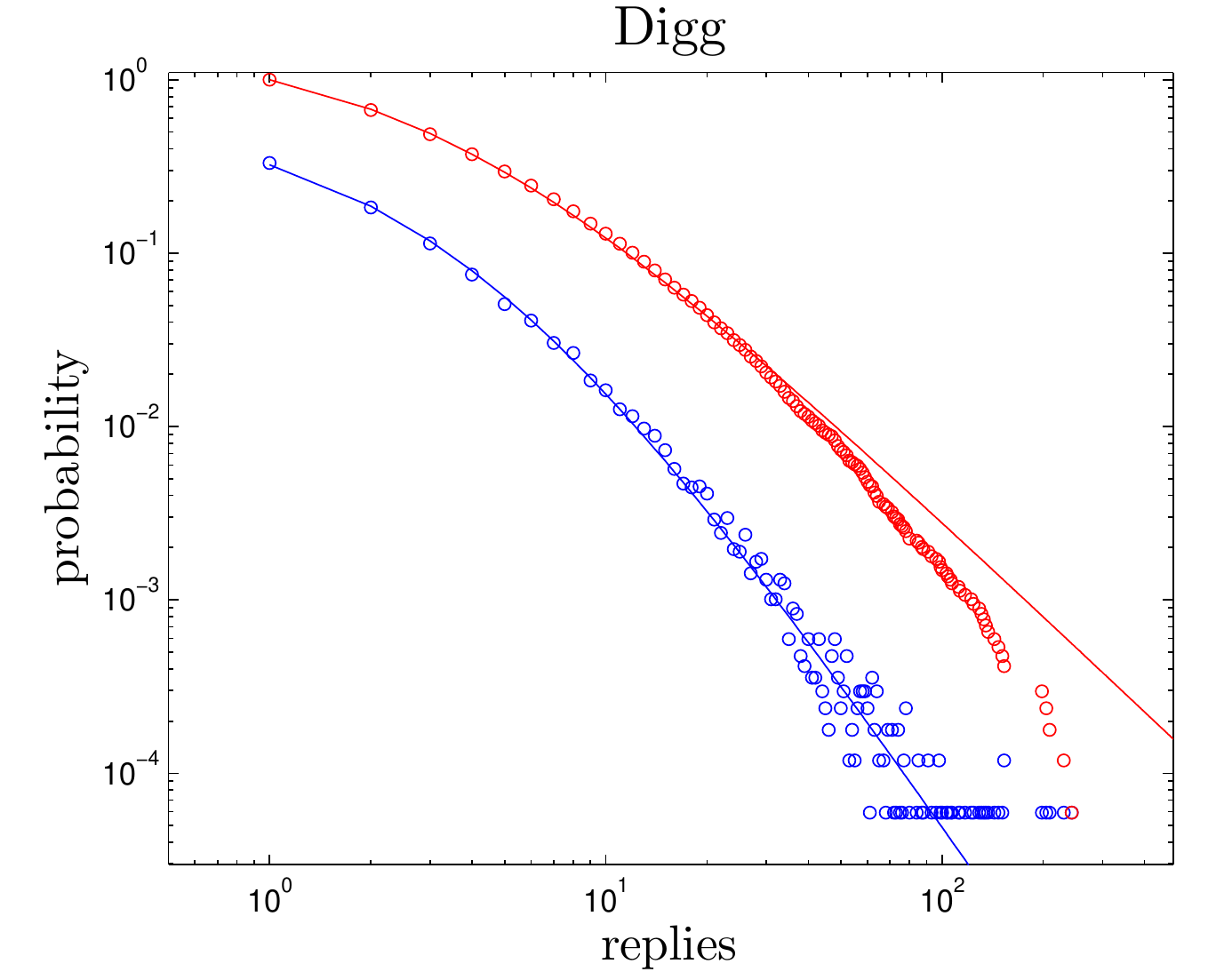}
\includegraphics[width=0.32\textwidth]{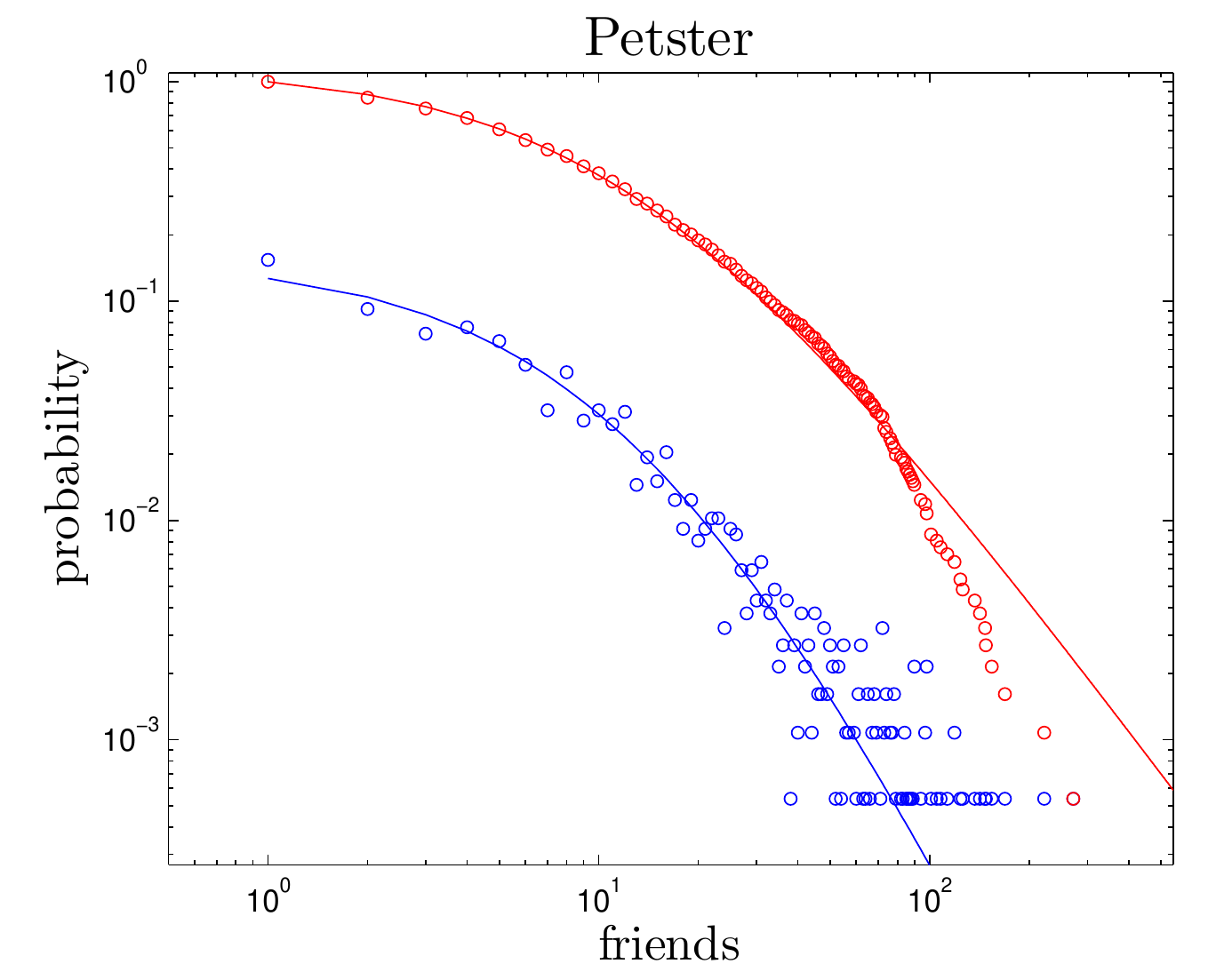}
\includegraphics[width=0.32\textwidth]{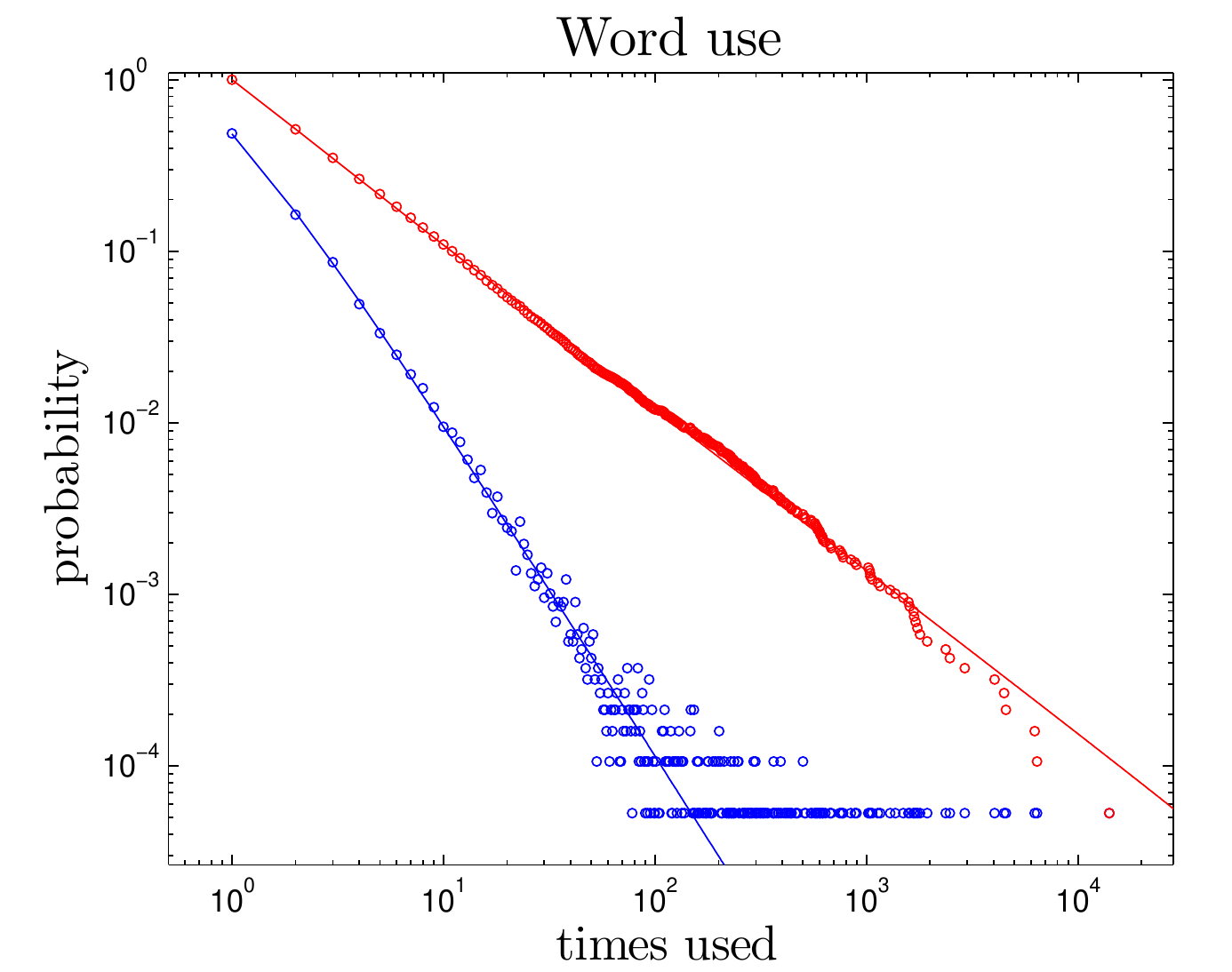}
}
\centerline{
\includegraphics[width=0.32\textwidth]{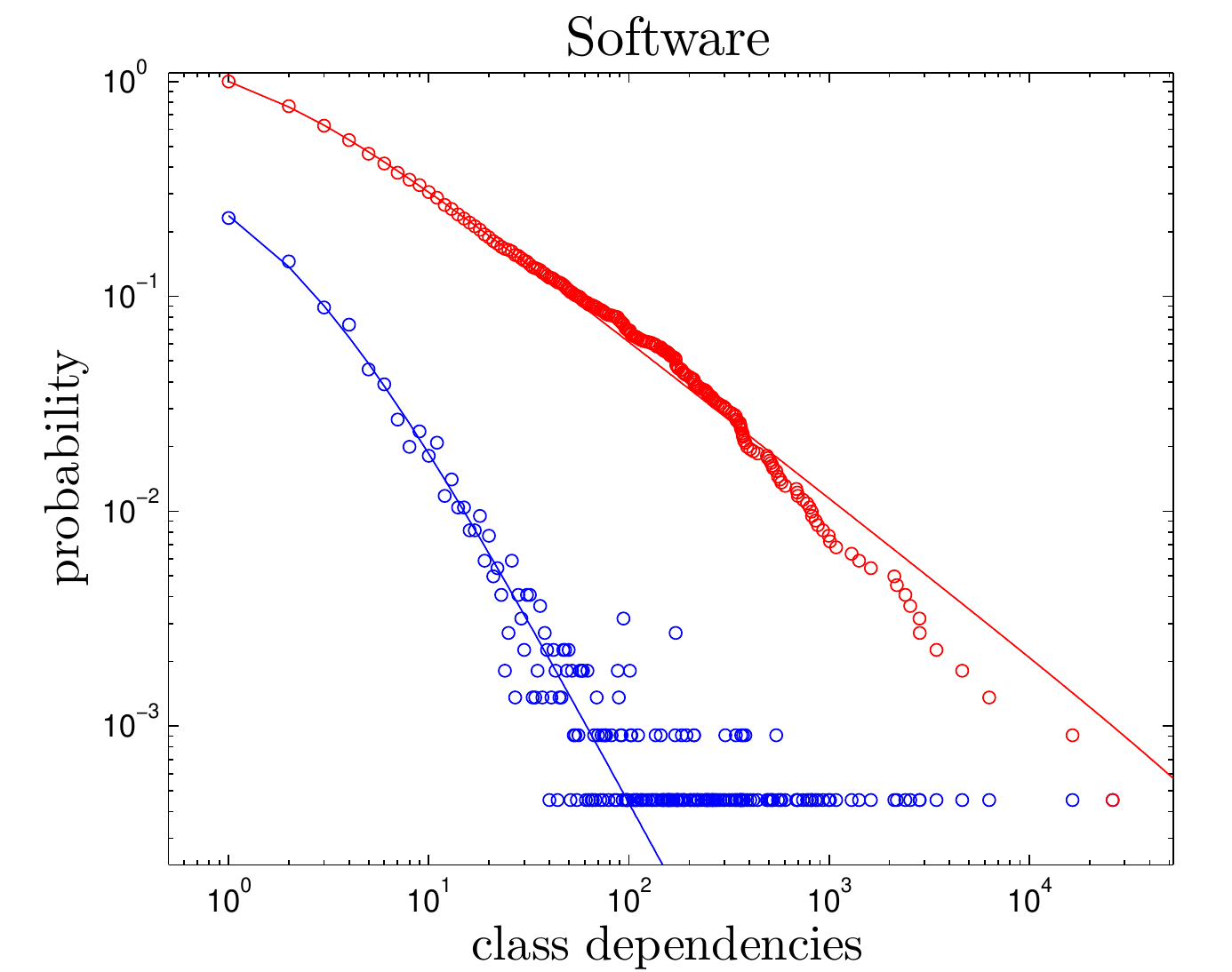}
}
\caption{Eq.~\ref{eq:partialsocialpk} gives good fits ($P > 0.05$; see SI for details) to 13 empirical distributions, with the values of $\mu^\circ$ and $k_0$ given in Table~1.  Points are empirical data, and lines represent best-fit distributions.  The probability $p_k$ of \emph{exactly} $k$ is shown in blue, and the probability of \emph{at least} $k$ (the complementary cumulative distribution, $\sum_{j=k}^{\infty} p_j$) is shown in red.  Descriptions and references for these datasets can be found in the SI.}
\label{fig:datafit}
\end{figure*}

In this language, the existing `community' is `paying down' some fraction of the joining costs for the next particle.  Mathematically, this communal cost-minus-benefit function can be expressed as
\beq
\label{eq:reduction}
\mu_k = \mu^\circ - \frac{k\mu_k}{k_0}.
\eeq
The quantity $\mu_k$ on the left side of Eq.~\ref{eq:reduction} is the total cost-minus-benefit when a particle joins a $k$-mer community.  The joining cost has two components, expressed on the right side: each joining event has an intrinsic cost $\mu^\circ$ that must be paid, and each joining event involves some discount that is provided by the community.  Because there are $k$ members of the existing community, the quantity $\mu_k /k_0$ is the discount given to a joiner by each existing community particle, where $k_0$ is a problem-specific parameter that characterizes how much of the joining cost burden is shouldered by each member of the community.  In the phone example, we assumed $k_0 = 1$.  The value of $k_0 = 1$ represents fully equal cost sharing between joiner and community member: each communal particle gives the joining particle a discount equal to what the joiner itself pays.  The opposite extreme limit is represented by $k_0 \rightarrow \infty$: in this case, the community gives no discount at all to the joining particle.

The idea of communal sharing of cost-minus-benefit is applicable to various domains.  It can express that one person is more likely to join a well-populated group on a social-networking site because the many existing links to it make it is easier to find (i.e., lower cost) and because its bigger hub offers the newcomer more relationships to other people (i.e., greater benefit).  Or, it can express that people prefer larger cities to smaller ones because of the greater benefits that accrue to the joiner in terms of jobs, services and entertainment. (In our terminology, a larger community `pays down' more of the cost-minus-benefit for the next immigrant to join.)  We use the terms `economy-of-scale' (EOS), or `communal', to refer to any system that can be described by a cost function such as Eq.~\ref{eq:reduction}, in which the community can be regarded as sharing in the joining costs, although other functional forms might also be of value for expressing EOS.

Rearranging Eq.~\ref{eq:reduction} gives $\mu_k = \mu^\circ k_0/(k+k_0)$.  The total cost-minus-benefit, $w_{k}$, of assembling a community of size $k$ is
\begin{align}\label{eq:Wkh}
w_{k} = \mu^\circ k_0 \sum_{j=1}^{k-1} \frac{1}{j+k_0} = \mu^\circ  k_0 \psi (k+k_0) - C,
\end{align}
where $\psi (k) = -\gamma + \sum_{j=1}^{k-1} j^{-1}$ is the digamma function ($\gamma = 0.5772...$ is Euler's constant), and the constant term $C = \mu^\circ k_0 \psi(k_0) + \mu^\circ$ will be absorbed into the normalization.

From this cost-minus-benefit expression (Eq.~\ref{eq:Wkh}), for a given $k_0$, we can now uniquely determine the probability distribution by maximizing the entropy.  Substituting Eq.~\ref{eq:Wkh} into Eq.~\ref{eq:sumQ} yields
\beq\label{eq:partialsocialpk}
p_k = \frac{\e^{-\mu^\circ k_0 \psi(k+k_0)}}{\displaystyle\sum_i \e^{-\mu^\circ k_0 \psi(i+k_0)}}.
\eeq

Eq.~\ref{eq:partialsocialpk} describes a broad class of distributions.  These distributions have a power-law tail {for large $k$}, with exponent $\mu^\circ k_0$, and a cross-over at $k=k_0$ from exponential to power-law.  To see this, expand $\psi (k+k_0)$ asymptotically and drop terms of order $1/k^2$.  This yields $w_{k} \sim \mu^\circ k_0 \ln \left(k+ k_0 -\tfrac{1}{2} \right)$, so Eq.~\ref{eq:partialsocialpk} obeys a power-law $p_k \sim \left( k+ k_0-\tfrac{1}{2}\right)^{-\mu^\circ k_0}$ for large $k$.  $p_k$ becomes a simple exponential in the limit of $k_0 \rightarrow \infty$ (zero cost sharing).  One quantitative measure of a distribution's position along the continuum from exponential-to-power-law is the value of its scaling exponent, $\mu^\circ k_0$.  A small exponent indicates that the system has extensive social sharing, thus power-law behavior.  As the exponent becomes large, the distribution approaches an exponential function. Eq.~\ref{eq:partialsocialpk} has a power-law scaling only when the cost of joining a community has a linear dependence on the community size.  The linear dependence arises because the joiner particle interacts identically with all other particles in the community.

What is the role of detailed balance in our modeling?  Figure~\ref{fig:schematic} shows no reverse arrows from $k$ to $k-1$.  The principle of maximum entropy can be regarded as a general way to infer distribution functions from limited information, irrespective of whether there is an underlying a kinetic model.  So, it poses no problem that some of our distributions, such as scientific citations, are not taken from `reversible' processes.

\section{Results\label{sc:results}}

\begin{figure}
\centerline{\includegraphics[width=0.5\textwidth]{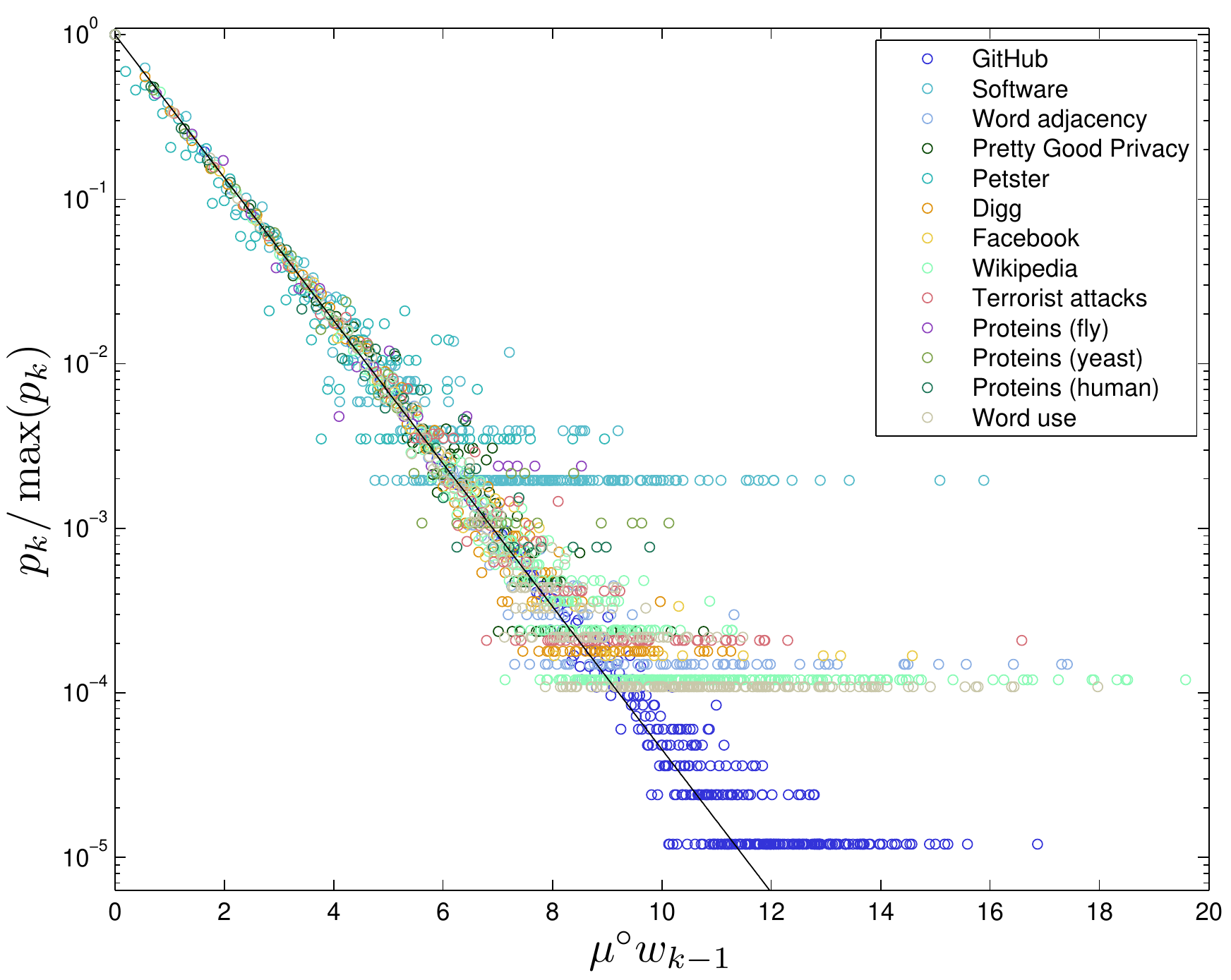}}
\caption{Eq.~\ref{eq:partialsocialpk} fitted to the 13 datasets in Table 1, plotted against the total cost to assemble a size $k$ community, $\mu^\circ w_{k-1}$.  Values of $\mu^\circ$ and $k_0$ are shown in Table 1.  The $y$-axis has been re-scaled by dividing by the maximum $p_k$, so that all curves begin at $p_k/\max(p_k)=1$.  All data sets are fit by the $\log y=-x$ line.  See Fig.~\ref{fig:datafit} for fits to individual datasets.}
\label{fig:linearized}
\end{figure}

Eq.~\ref{eq:partialsocialpk} and Fig.~\ref{fig:datafit} show the central results of this paper.  Consider three types of plots.  On the one hand, exponential functions can be seen in data by plotting $\log p_k$ vs $k$.  Or, power-law functions are seen by plotting $\log p_k$ vs $\log k$.  Here, we find that plotting $\log p_k$ vs a digamma function provides a universal fit to several disparate experimental data sets over their full distributions (Fig.~\ref{fig:linearized}).  Fig.~\ref{fig:datafit} shows fits of Eq.~\ref{eq:partialsocialpk} to 13 datasets, using $\mu^\circ$ and $k_0$ as fitting parameters that are determined by a maximum-likelihood procedure.  (See SI for dataset and goodness-of-fit test details.) $\mu^\circ$ and $k_0$ characterize the intrinsic cost of joining any cluster, and the communal contribution to sharing that cost, respectively.

Rare events are less rare under fat-tailed distributions than under exponential distributions.  For dynamical systems, the risk of such events can be quantified by the \emph{coefficient of variation} (CV), defined as the ratio of the standard deviation $\sigma_k$ to the mean $\la k \ra$. For equilibrium/steady state systems, the CV quantifies the spread of a probability distribution, and is determined by the power-law exponent, $\mu^\circ k_0$.  Systems with small scaling exponents ($\mu^\circ k_0 \le 3$) experience an unbounded, power-law growth of their CV as the system size $N$ becomes large, $\sigma_k/\la k \ra \sim N^\beta$.  This growth is particularly rapid in systems with $1.8 < \mu^\circ k_0 < 2.2$, because the average community size $\la k \ra$ diverges at $\mu^\circ k_0 = 2$.  For these systems, $\beta = 1/2$ is observed.  Several of our datasets fall into this `high-risk' category, such as the number of deaths due to terrorist attacks (Table~1).

\begin{table}[hbdp]
\caption{Fitting parameters and statistics.}
\begin{center}
\begin{tabular}{l | l | l | l | l | l | l}
\hline
Data set & $\mu^\circ$ & $k_0$  & $\la k \ra$ & $N$ & $\mu^\circ k_0$ & $P$ \\
\hline
GitHub & 9(1) & 0.21(2) & 3.642 & 120,866 & 2(2) & 0.78\\
Wikipedia & 1.5(1) & 1.3(1) & 25.418 & 21,607 & 1.9(1) & 0.79\\
PGP & 1(1) & 2.6(2) & 4.558 & 10,680 & 2.6(3) & 0.16\\
Word adjacency & 3.6(4) & 0.6(1) & 5.243 & 11,018 & 2.1(3) & 0.09\\
Terrorists & 2.1(2) & 1(1) & 4.346 & 9,101 & 2.2(3) & 0.38\\
Facebook wall & 1.6(1) & 2.3(3) & 2.128 & 10,082 & 3.6(5) & 0.99\\
Proteins, fly & 0.9(2) & 5(2) & 2.527 & 878 & 5(2) & 0.89\\
Proteins, yeast & 0.9(1) & 4(1) & 3.404 & 2,170 & 3(1) & 0.48\\
Proteins, human & 0.8(1) & 4(1) & 3.391 & 3,165 & 4(1) & 0.52\\
Digg & 0.68(3) & 4.2(3) & 5.202 & 16,844 & 2.8(2) & 0.05\\
Petster & 0.21(3) & 15(3) & 13.492 & 1,858 & 3(1) & 0.08\\
Word use & 2.3(1) & 0.8(1) & 11.137 & 18,855 & 1.9(2) & 0.56\\
Software & 0.8(1) & 2.1(3) & 62.82 & 2,208 & 1.7(3) & 0.69\\
\hline
\end{tabular}
\end{center}
\label{tab:params}
\end{table}

\section{Discussion\label{sc:disc}}

We have expressed a range of probability distributions in terms of a generalized energy-like cost function. In particular, we have considered types of costs that can be subject to economies of scale, which we have also called `community discounts'.  We maximize the Boltzmann-Gibbs-Shannon entropy, subject to such cost-minus-benefit functions.  This procedure predicts probability distributions that are exponential functions of a digamma function.  Such a distribution function has a power-law tail, but reduces to a Boltzmann distribution in the absence of EOS.  This function gives good fits to distributions ranging from scientific citations and patents, to protein-protein interactions, to friendship networks, and to weblinks and terrorist networks -- over their full distributions, not just in their tails.  

Framed in this way, each new `joiner particle' must pay an intrinsic buy-in cost to join a community, but that cost may be reduced by a communal discount (an economy of scale).Here, we discuss a few points. First, both exponential and power-law distributions are ubiquitous. How can we rationalize this? One perspective is given by switching viewpoint from probabilities to their logarithms, which are commonly expressed in a language of dimensionless cost functions, such as energy$/RT$.  There are many forms of energy (gravitational, magnetic, electrostatic, springs, interatomic 
interactions, etc).  The ubiquity of the exponential distribution can be seen in terms of the diversity and interchangeability of energies.  

A broad swath of physics problems can be expressed in terms of the different types of energy and their ability to combine, add or exchange with each other in various ways. Here, we indicate that non-exponential distributions too, can be expressed in a language of costs, particularly those that are shared and are subject to economies of scale.  Second, where do we expect exponentials vs. power-laws?  What sets Eq.~\ref{eq:reduction} apart from typical energy functions in physical systems is that EOS costs are both independent of distance and long-ranged (the joiner particle interacts with all particles in given community).  Consequently, when the system size becomes large, due to the absence of a correlation length-scale, the energy of the system does not increase linearly with system size giving rise to a non-extensive energy function. This view is consistent with the appearance of power-laws in critical phenomena, where interactions are effectively long-ranged.

Third, interestingly, the concept of cost-minus-benefit in Eq.~\ref{eq:reduction} can be further generalized, also leading to either Gaussian or stretched-exponential distributions.  A Gaussian distribution results when the cost-minus-benefit function grows linearly with cluster size, $\mu_k \sim k$; this would arise if the joiner particle were to pay a tax to each member of a community.  This leads to a total cost of $w_k \sim k^2$ (see Eq.~\ref{eq:cumu}).  These would be `hostile' communities, leading to mostly very small communities and few large ones, since a Gaussian function drops off even faster with $k$ than an exponential does.  An example would be a Coulombic particle of charge $q$ joining a community of $k$ other such charged particles, as in the Born model of ion hydration~\cite{born1920volumes}.  A stretched-exponential distribution can arise if the joiner particle instead pays a tax to only a subset of the community.  For example, in a charged sphere with strong shielding, if only the particles at the sphere's surface interact with the joiner particle, then $\mu_k \sim k^{2/3}$ and $w_k \sim k^{5/3}$, leading to a stretched-exponential distribution. In these situations, EOS can affect the community-size distribution not only through cost sharing but also through the topology of interactions.

Finally, we reiterate a matter of principle.  On the one hand, non-exponential distributions could be derived by using a non-extensive entropy-like quantity, such as those of Tsallis, combined with an extensive energy-like quantity.  Here, instead, our derivation is based on using the Boltzmann-Gibbs-Shannon entropy combined with a non-extensive energy-like quantity.  We favor the latter because it is consistent with the foundational premises of Shore and Johnson~\cite{Shore_1980}.  In short, in the absence of energies or costs, the BGS entropy alone predicts a uniform distribution; any other alternative would introduce bias and structure into $p_k$ that is not warranted by the data.  Models based on non-extensive entropies, on the other hand, intrinsically prefer larger clusters, but without any basis to justify them.  The present treatment invokes the same nature of randomness as when physical particles populate energy levels.  The present work provides a cost-like language for expressing various different types of probability distribution functions.

\begin{acknowledgments}
We thank A.~de~Graff,  H.~Ge, D.~Farrell, K.~Ghosh, S.~Maslov, and C.~Shalizi for helpful discussions, and K.~Sneppen, M.S.~Shell, and H.~Qian for comments on our manuscript.  J.P. thanks the U.S.~Department of Defense for financial support from a National Defense Science and Engineering Graduate Fellowship.  J.P. and K.D. thank the NSF and Laufer Center for support.  P.D. acknowledges Department of Energy grant PM-031 from the Office of Biological Research.
\end{acknowledgments}

\cleardoublepage

\setcounter{figure}{0}
\setcounter{table}{0}
\setcounter{page}{1}
\makeatletter
\renewcommand{\thefigure}{S\@arabic\c@figure}
\makeatletter
\renewcommand{\thetable}{S\@arabic\c@table}

\begin{figure}[b]
\centering{\includegraphics[width=0.45\textwidth]{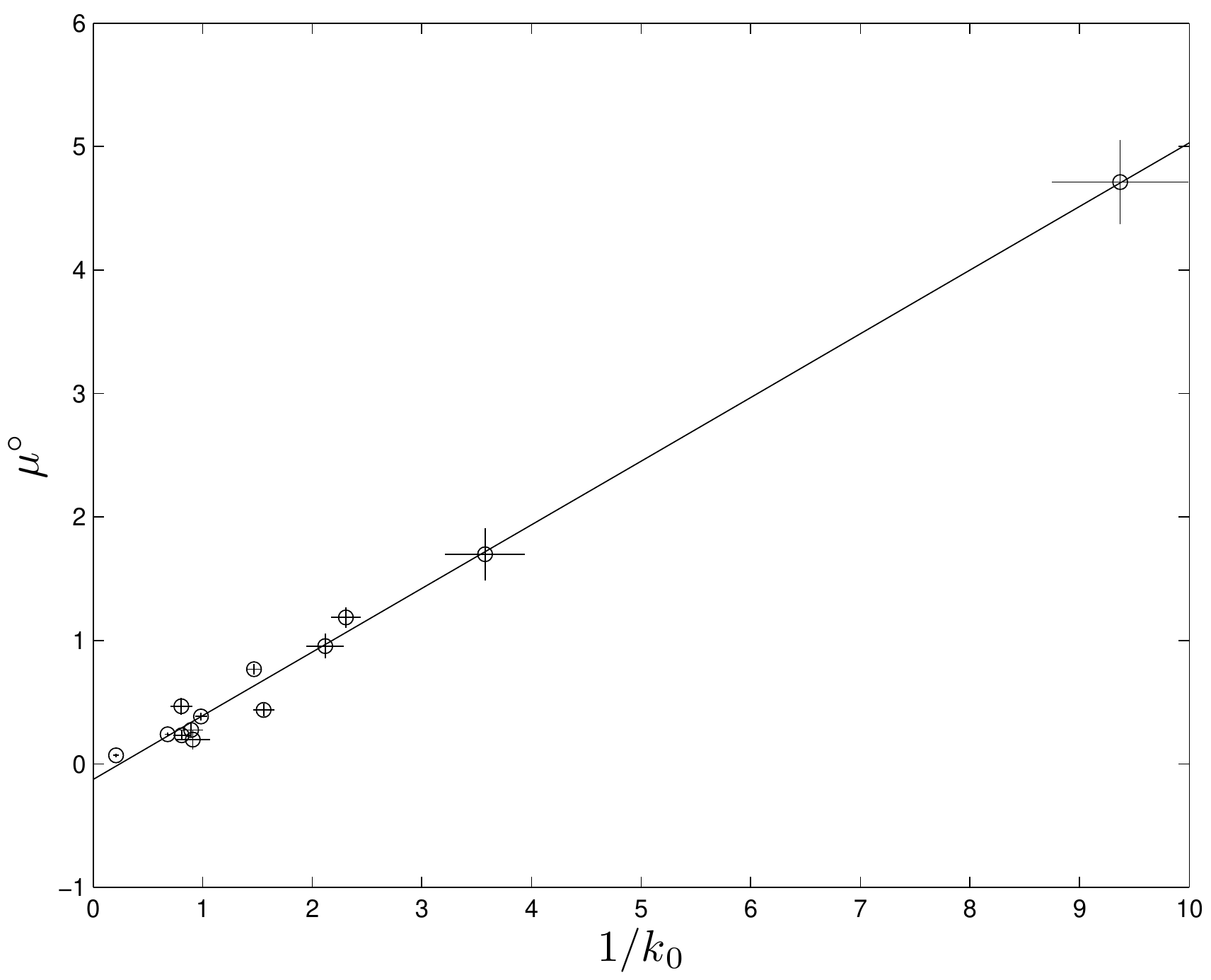}}
\caption{$\mu^\circ$ plotted against $1/k_0$ for the 13 data sets listed in Table \ref{tab:params}.  Error bars are 95\% confidence intervals.  Linear regression is shown as a solid line, $\mu^\circ =  0.516 k_0^{-1} - 0.125$ ($R^2 = 0.991$).  Each point on this plot represents an empirical data set.}
\label{fig:mu_s}
\end{figure}

\begin{appendix}
\section{Datasets and fitting}

In this paper, we have two aims: (i) {to use} Max Ent to identify a general functional form for situations involving cost-minus-benefit constraints, and (ii) to fit data over a broad range of contexts.  For (i), Max Ent predicts an exponential of a digamma function, provided that $k_0$ is known.  It also shows how to compute a full distribution if we are given a single quantity, $\langle w \rangle$.  For (ii), our aim is just to do simple curve-fitting of data, given the mathematical form from (i).  In this case, we have no microscopic model for $k_0$, so we know neither $k_0$ nor $\mu^\circ$.  In this case, our objective is not to find the full distribution from $\langle w \rangle$.  Rather, for (ii), we are given the full distribution function, and our objective is to find the best values of $k_0$ and $\mu^\circ$ that fit it.

Our fitting procedure is as follows.  We use the maximum likelihood estimation function \texttt{mle} function in Matlab, with probability distribution specified by Eq.~\ref{eq:partialsocialpk}.  Table~\ref{tab:params} shows estimated $\mu^\circ$ and $k_0$ values, with 95\% confidence intervals.  We calculate goodness-of-fit $P$-values using the Monte Carlo simulation procedure (based on the Kolmogorov-Smirnov test) described in~\cite{Clauset_2009}.  The 13 datasets shown here are above the $P = 0.05$ statistical significance threshold proposed in~\cite{Clauset_2009}.  $P$-values shown in Table~S1 are based on 1000 simulations for each dataset.

We fit Eq.~\ref{eq:partialsocialpk} to 13 data sets:
\begin{itemize}
\item Project membership on the social coding website GitHub (downloaded from \texttt{konect.uni-koblenz.de})
\item Edits made by users of the English-language Wikipedia \cite{download.wikimedia.org}
\item Interactions between users of the Pretty Good Privacy (PGP) secure data transfer algorithm \cite{Boguna_2004}
\item Words occurring immediately after one another in a Spanish book (downloaded from \texttt{konect.uni-koblenz.de})
\item Deaths resulting from terrorist attacks from February 1968 to June 2006 \cite{Clauset_2007}
\item Wall posts by users to their walls on the social-networking website Facebook, from a 2009 crawl of New Orleans Facebook \cite{Viswanath_2009}
\item Pairwise, physical protein-protein interactions (PPI) of proteins detected in small-scale PPI network data, in yeast (\emph{Saccharomyces cerevisiae}), fruit flies (\emph{Drosophila melanogaster}), and humans (\emph{Homo sapiens}) \cite{Patil_2011}
\item Replies between users of the social news website Digg \cite{b565}
\item Friendships between users of the Petster social networking site Hamsterster (downloaded from \texttt{konect.uni-koblenz.de})
\item Occurrences of unique words in the novel \emph{Moby Dick} \cite{Newman_2005}
\item Class-class dependencies in the software libraries JUNG and javax (downloaded from \texttt{konect.uni-koblenz.de})
\end{itemize}
An overlay of all fits and datasets is shown in Fig.~\ref{fig:linearized}.  Individual parameters and fits are shown in Table~\ref{tab:params} and Fig.~\ref{fig:datafit}.

\end{appendix}

\end{document}